# OptoCENTAL: a standardised, bench-testing platform based on phantoms for validating optical systems aimed at clinical monitoring of the placenta

Luca Giannoni,[a*] Uzair Hakim,[a] Fréderic Lange,[a] Musa Talati,[a] Darshana Gopal,[a] Angelos Artemiou,[a] Niccole Ranaei-Zamani,[b] Subhabrata Mitra,[b] Ilias Tachtsidis[a]

[a] University College London, Department of Medical Physics and Biomedical Engineering, London, UK
[b] University College London, Neonatology, EGA Institute for Women's Health, London, UK

**Abstract**

**Significance:** Optical imaging and spectroscopy solutions, such as near-infrared spectroscopy (NIRS) and diffuse optical tomography (DOT), have the potential to provide compact, bedside monitoring of the placenta in the clinic, thanks to recent advancements in miniaturisation and wireless wearability. This would provide neonatologist with continuous assessment of the pregnancy status in real-time, as well as tools to possibly predict delivery outcomes.

**Aim:** Numerous challenges are associated with validating the performances of any optical system in providing the above goals, particularly the required depth sensitivity and accuracy to target the placenta *in vivo*. Currently, no testing procedure is available that can verify in a thorough and comparable manner the capability of a given photonic system to be successful in the clinics at placental monitoring.

**Approach:** We present here OptoCENTAL, a standardized platform based on multiple optical phantoms, from digital, through solid to liquid, for a comprehensive bench-testing, characterisation and validation of any photonics solution and instrumentation that aims at *in vivo*, clinical monitoring of the human placenta.

**Results:** Exemplary applications of the OptoCENTAL platform on different types of optical systems, from wearable, continuous-wave devices to broadband and time-domain NIRS systems, demonstrate the flexibility of its procedures to be implemented with any setup, allowing users to compare performances across different solutions. The results also show the capability of OptoCENTAL to provide quantitative assessment of the major features required by any photonic solution for providing effective and efficient monitoring of the placenta, including basic instrument performances, quantification of monitoring accuracy, as well as depth sensitivity.

**Conclusions:** OptoCENTAL represent the first-of-a-kind effort in standardising bench-testing and validation of optical imaging and spectroscopy methods in the framework of placental clinical applications, further advancing the translation of such modalities into the hospitals, as well as towards future certification and commercialisation of such technologies.

*Luca Giannoni, E-mail: l.giannoni@ucl.ac.uk

## 1 Introduction

Bedside, non-invasive and continuous monitoring of the human placenta during gestation is of utmost importance in the clinics for assessing in real-time the status of the pregnancy (and, indirectly, of the foetus) and potentially predicting any complications that may occur during delivery, such as reduced or lack of oxygenation to the foetus or foetal metabolic disruptions, to mention a few critical instances. Some of these conditions represent, or are strictly connected to, the leading causes in adverse pregnancy outcomes, including perinatal mortality, stillbirths, or even long-term morbidity conditions affecting the newborn[1,2].

The physiological status of the placenta thus can act as an essential analytical indicator of the positive or negative progression of a pregnancy, and its continuous screening could allow physicians to provide more direct and prompt intervention in case any clinical emergency manifests. Furthermore, recent studies on placental physiology have now started correlating with high accuracy the status of the placenta during gestation with adverse neonatal outcomes[3]. Therefore, easier access to information on the physiological and/or pathophysiological status of the placenta in real-time could have the outstanding benefit of allowing neonatologists to predict in advance any possible risks of negative and unfavourable delivery, as well as any undesired consequences for the newborn's health[4].

## 1.1 Optical monitoring of the placenta

Within this main clinical framework, optical imaging and spectroscopy technologies, such as near infrared spectroscopy (NIRS) and diffuse optical tomography (DOT), represent promising, quantitative tools[5–8] for bedside, streamlined and continuous physiological monitoring of the placenta non-invasively, in a cost-effective, user-friendly and compact fashion (that can be easily integrated in hospital settings). Such types of monitoring modalities are based on the continuous measurement of placental haemodynamics and oxygenation, that are inferred from the intrinsic optical signatures generated from the absorption of near infrared (NIR) light by oxygenated ($HbO_2$) and deoxygenated (HHb) haemoglobin[9,10]. NIR light in the range from 650 to 100 nm, specifically, has the advantage of providing a much deeper sampling power in human tissues, compared to visible one, thus opening to the capability to travel within the abdomen and reach the placenta underneath[11,12].

In addition to haemodynamic monitoring, information regarding the metabolism of the investigated tissue can also be quantified using multi-wavelength or broadband NIRS approaches, that employs a large number (tens to hundreds) of wavelengths for targeting the wide NIR absorption features (780-900 nm) from the redox states of cytochrome-c-oxidase (CCO)[13,14]. CCO is an enzyme in the electron transport chain of the mitochondria that is responsible for more than 95% of adenosine triphosphate (ATP) production, thus can act as an optical biomarker to metabolic activity and related changes in mitochondrial functions, which are all crucial during gestation and foetal development[15].

Finally, to further enhance translation of optical technologies towards the clinic and to make them suitable for ergonomical applications on patients, recent advancements have also further improved the compactness and transportability of photonics instrumentation for placental monitoring, paving the way for more user-friendly, wearable and wireless medical devices [12,16].

*1.2 Current challenges in clinical applications*

Several challenges in sensitivity and accuracy are indeed present when using optical modalities for monitoring the placenta *in vivo*, predominantly related to its depth, geometry and position, as well as the presence of several abdominal overlayers of various thicknesses above it.

The human placenta can range from 1 to 4 cm deep within the body from the top of the skin surface, according to the gestational stage, as well as inter-subject variability due to different reasons, including the body type of the parent[17]. Placentas can also present as anterior, when at the front of the uterus, or posterior, when at the back. The overlayers above the placenta can also vary significantly depending on the position of the placenta and other anatomical factors: the main layers, on average from 1 to 2 cm in total thickness, are composed primarily of muscle (abdominal and uterine muscles, mainly), skin and adipose tissue, whilst water-based liquor layers can also manifest, and bone can sometimes be present (the spine and the pelvis) in case posterior placentas. Structural analysis of the placenta and overlayers is typically performed via ultrasound scans as the standard clinical procedure[18]. All the above aspects undermine the capability of NIR light to penetrate deep enough to reach the placenta, as well as to provide enough signal-to-noise ratio in the detected intensity to be used to estimate valuable physiological information.

*1.3 The need for standardised testing and validation*

Given the existing challenges in clinical monitoring of the placenta, a suitable, systematic and comprehensive validation of the performance of any optical system aiming at recovering meaningful and accurate information from oxygenation and metabolism of the placenta is essential before any clinical translation can be taken into consideration: this is crucial for robustly and confidently establishing whether a given photonics-based solution is capable or not to effectively sample at the required depth in the human body and with adequate precision and sensitivity to the main biomarkers of interest, such as $HbO_2$, HHb and CCO. Albeit few attempts to standardised testing procedures exist in the NIRS community for applications on the brain, such as the MEDPHOT[19] and the NeurOPT[20] protocols, currently no standardised and extensive set of procedures is available specifically for the placenta, in either the clinical nor the technical community, that can provide efficient bench-testing, characterisation and validation of optical instrumentation specifically for the task of placental monitoring at bedside, and with the ultimate goal of ensuring reliability in the application of any photonics device in the neonatal units.

For this purpose, we present here the development of OptoCENTAL: a standardised platform for bench-testing and validation of any optical system aimed at placental monitoring, from continuous-wave (CW) to time-domain (TD) NIRS devices, through multi-wavelength, broadband (bNIRS) and even wearable sensors. The OptoCENTAL platform is based on a repeatable set of quantitative assessment procedures involving optical phantoms, including digital and physical ones, from solid, to liquid to hybrid and multilayered[21–24]. Optical phantoms have become a gold standard in the biomedical diffuse optics community, for their capability to act as controllable and realistic proxies of the human tissue. For OptoCENTAL, custom-made phantoms have been engineered to mimic the optical properties, structure and geometry of the human abdomen and placenta, to methodically and systematically test any light-based instrumentation. Examples of its inner working and application to existing photonics instruments engineering for placental monitoring are presented here, including two wearable, multi-wavelength CW-NIRS devices, a wearable bNIRS setup, as well as a multi-wavelength TD-NIRS system.

The purpose of OptoCENTAL is to ultimately lead users to a complete, reliable, fast and cross-comparable assessment of the performances and capability on any photonics setup aimed at clinical applications for monitoring the human placenta *in vivo*.

## 2 The OptoCENTAL platform

OptoCENTAL, the standardised, bench-testing platform for placental monitoring optical instruments here presented, is composed of four distinct protocols utilising different types and forms of custom-made and easily reproducible optical phantoms. Each protocol covers a specific set of features concerning both the tested photonic equipment and the biological target scenario (i.e., the placenta), which can be tailored to adapt to any type of instrumentation, as well as by considering the physiological specificity of the application (by introducing *ad hoc* information regarding human placentas). All the protocols are envisioned to be easily completed in a systematic and rapid way in less than a week of measurements and data analysis. The following is a brief overview of each of the four protocols of OptoCENTAL:

- PROTOCOL 0: A computational depth sensitivity assessment protocol based on a multi-layered Monte Carlo simulator that uses a realistic digital phantom of a real placenta, generated from medical ultrasound scans routinely performed in the clinics, as well as a digital twin of the tested optical instrumentation for placental monitoring.
- PROTOCOL 1: A basic instrument assessment protocol of the standard performances of any photonics device for placental monitoring, including the quantification of its signal-to-noise ratio (SNR), noise/variance, linearity, stability and reliability/repeatability. PROTOCOL 1 is based on a set of customised solid, homogenous phantoms with selected and fixed, known optical properties.
- PROTOCOL 2: A physiological monitoring assessment protocol of the capability and accuracy of any photonics solutions for placental monitoring in reconstructing continuous changes in tissue haemodynamics and metabolism (without crosstalk). PROTOCOL 2 is based on a homogeneous, dynamic liquid phantom recipe with biological contrast agents (water, intralipid, blood and yeast).
- PROTOCOL 3: A depth sensitivity assessment of the capability and accuracy of any photonics system for placental monitoring in discriminating signals deep within abdominal tissue and specifically from the placenta. PROTOCOL 3 is based on a multilayered, hybrid (solid and liquid), dynamic phantom.

*2.1 PROTOCOL 0: Monte Carlo digital phantom and instrument simulator*

Monte Carlo simulations for PROTOCOL 0 were carried out using the Monte Carlo eXtreme (MCX) simulation package with Python[25] and performed on an NVIDIA Clara system (8 core

Carmel ARM CPU, 32GB LPDDR4 RAM, NVidia RTX6000 24GB GPU)[26]. A 200-mm³ volume was created for each simulation, with 15e⁹ photons released into the volume. To replicate the general abdomen physiology, the simulation volume was discretised into four layers: skin, muscle, adipose and placenta. Ultrasound scans were acquired on 268 healthy participants during maternal gestation at the Institute for Women's Health in the University College London Hospital (UK), at different gestational stages (ranging from 24 to 40 weeks)[27]. The ultrasound scans were then analysed by clinicians to define the depths of the skin, muscle and adipose layers. These depths were then used to construct realistic simulation volumes for PROTOCOL 0, where examples are provided in Sec. 3.1. The remainder of the 200-mm depth was defined as placenta.

Optical properties for the skin layer were calculated using the fraction of melanosome associated to each real patient data. Clinicians defined patient skin colour using the Fitzpatrick scale, where skin colour is given a number from 1-6 based on hue tone, with 1 being the lightest. From these, associated fraction of melanosome $F_{mel}$ was ascribed based on the values from Karsten *et al*[28]. This relates how much melanosome is present in each Fitzpatrick score. These associations are reported in Table 1.

**Table 1** Volume fractions of melanosome associated to Fitzpatrick score, as used in PROTOCOL 0.

| Fitzpatrick score | Volume fractions of melanosome |
|---|---|
| 1 | 0.0255 |
| 2 | 0.0255 |
| 3 | 0.0902 |
| 4 | 0.155 |
| 5 | 0.155 |
| 6 | 0.3055 |

The absorption coefficient $\mu_{a,skin}$ for the skin is then defined as the sum of the contributions from both the absorption coefficient of the dermis $\mu_{a,dermis}$ and that of the epidermis $\mu_{a,epid}$. The melanosome content is contained within the epidermis, whose absorption coefficient $\mu_{a,epid}$ is calculated using the same method as in Jacques et al[29], following the formula:

$$\mu_{a,epid} = F_{mel} \cdot \mu_{a,mel} + (1 - F_{mel}) \mu_{a,skin(baseline)} \qquad (1)$$

$\mu_{a,skin(baseline)}$ is taken from Oltulu et al[30].

The absorption coefficient of melanin $\mu_{a,mel}$ is obtained from and described as:

$$\mu_{a,mel} = 510 \left(\frac{\lambda}{500\text{nm}}\right)^{-3} \qquad (2)$$

The absorption coefficient for the dermis $\mu_{a,dermis}$ is then computed using the following formula from Jacques et al[29]:

$$\mu_{a,dermis} = (1 - F_b)\mu_{a,skin(baseline)} + \ln 10 \cdot HbT_{dermis}[StO_{2,dermis} \cdot \varepsilon_{HbO_2} + \qquad (3)$$
$$+ (1 - StO_{2,dermis}) \cdot \varepsilon_{HHb}]$$

Where $F_b$ is the volume fraction of blood in the dermis, $HbT_{dermis}$ is the total haemoglobin in the dermis layer of the skin, $StO_{2,dermis}$ is the oxygen saturation in the dermis layer, and $\varepsilon_{HbO_2}$ and $\varepsilon_{HHb}$ are the extinction coefficients for $HbO_2$ and $HHb$, respectively. All these parameters were obtained from Jacques et al[29].

Following the computation of the absorption coefficients of the dermis and epidermis, these are combined to obtain the absorption coefficient of the skin. This was done with the consideration of the fractional contribution of each layer to the total skin. Using the formula from Oltulu et al[30], the fraction of dermis $F_{dermis}$ and epidermis $F_{epid}$ was computed based on the thicknesses of the dermis and epidermis, taken as 4550 μm and 127.2 μm respectively. Thus, the final absorption coefficient for the skin is computed as:

$$\mu_{a,skin} = F_{dermis} \cdot \mu_{a,dermis} + F_{epid} \cdot \mu_{a,epid} \tag{4}$$

The absorption coefficients for $\mu_{a,n}$ the remaining $n$ tissues (muscle, adipose, placenta) were calculated based on:

$$\mu_{a,n} = F_{fat} \cdot \mu_{a,fat} + F_{H_2O} \cdot \mu_{a,H_2O} + \ln 10 \cdot C_{HbT}[StO_2 \cdot \varepsilon_{HbO_2} + \\ +(1 - StO_2) \cdot \varepsilon_{HHb}] \tag{5}$$

Where $F_{fat}$ is the volume fraction of fat in the tissue, $\mu_{a,fat}$ is the absorption coefficient of fat, $F_{H_2O}$ is the volume fraction of water in the tissue, $\mu_{a,H_2O}$ is the absorption coefficient of water, and $C_{HbT}$ is the total concentration of haemoglobin HbT in the tissue. The contribution of each value to a given tissue's absorption coefficient is given by Table 2.

Table 2 Volume content contributions of water and fat in each type of tissues for the calculation of the absorption coefficients used in PROTOCOL 0.

| Tissue | $F_{H_2O}$ (%) | $F_{fat}$ (%) |
|---|---|---|
| Skin | 0 | 0 |
| Adipose | 80 | 20 |
| Muscle | 76 | 0 |
| Placenta | 85 | 0 |

As can be seen from Eq. 5, the absorption coefficient of each tissue is also impacted by their content of HbT and their $StO_2$. For the adipose layer, these values are not relevant and are set to 0, because it is assumed that no significant blood flow is present. For muscle and placenta, these values are user defined and were set at a constant of $C_{HbT}$ equal to 50 µMol for both, whilst $StO_2$ was set to 60% for muscle and 80% for placenta.

Sources and detectors were positioned according to the optical configuration of the device being simulated, as described in specific details in Sec. 3.1.

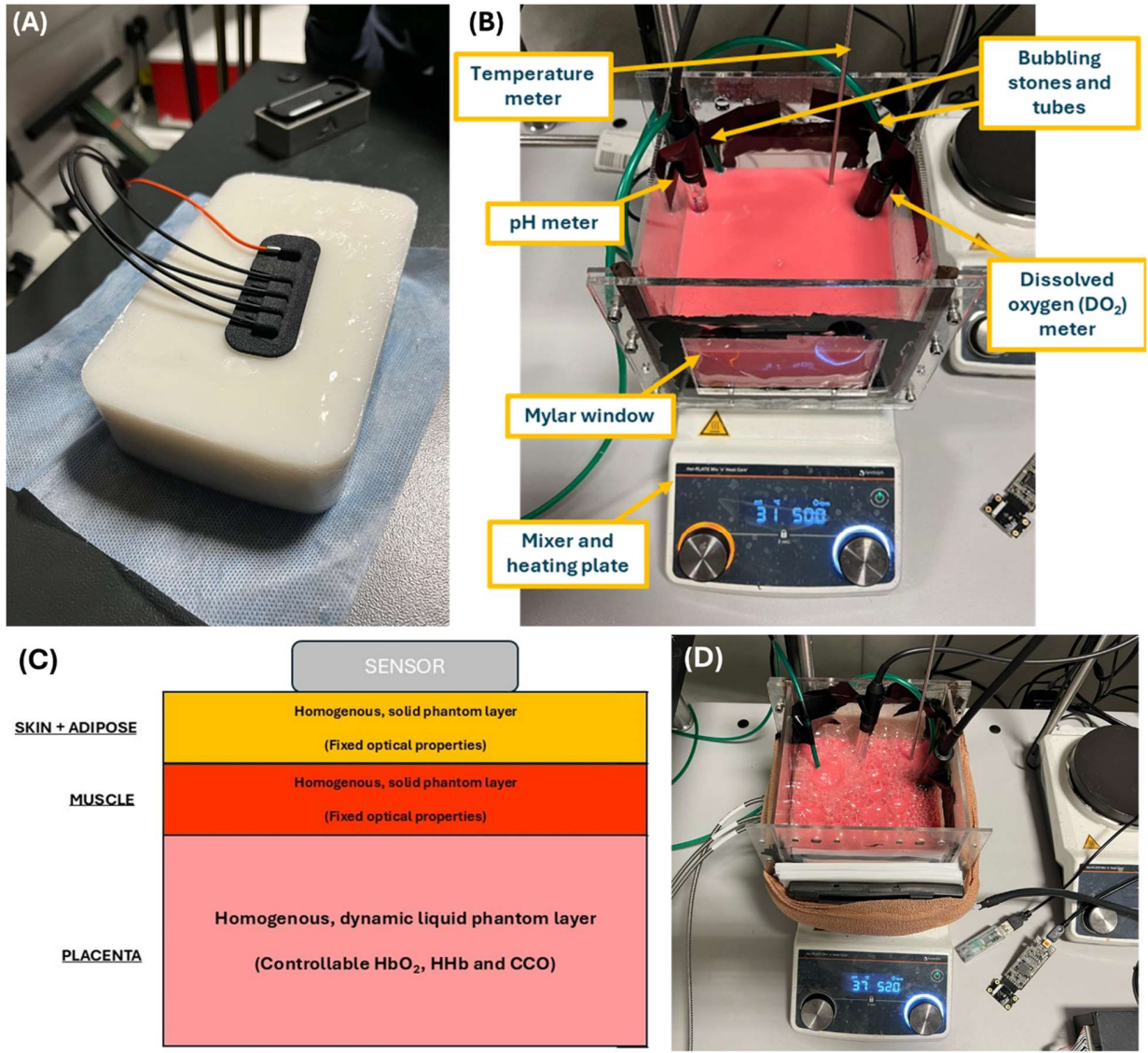

**Fig. 1** (A) Example of one of the custom-made, homogenous, solid phantoms used in PROTOCOL 1 and of the configuration for the characterisation of its optical properties, using MAESTROS II[4]; (B) Picture of the homogenous, liquid phantom used in PROTOCOL 2 and its housing, with all its main components indicated, including controllers and probes; (C) Diagram of the multi-layered structure of the hybrid phantom sued in PROTOCOL 3; (D) Picture of the multi-layered, hybrid phantom used in PROTOCOL 3, with an example of wearable NIRS device attached.

## 2.2 PROTOCOL 1: Basic instrument assessment with solid phantoms

PROTOCOL 1 is based on analogous testing procedures as the well-established and standardised MEDPHOT protocol[19] in the NIRS community. Its main purpose is to provide basic instrument characterisation, testing for the following technical aspects of any optical device: (1) SNR, i.e. overall quantification of the signal quality over the corresponding baseline noise (across the whole spectral operational range); (2) noise/variance in the signal, i.e., the sensitivity and influence of the device to statistical fluctuations in their readings over time; (3) linearity in the measurements of phantoms with different optical absorption, to check whether the sensors can follow changes in a given parameter without distortions; (4) stability of the signal across a repeated measurements over short periods of time (i.e., 1 hour); (5) repeatability of the output of the measurements under the same experimental conditions over three consecutive days.

For the above purposes, PROTOCOL 1 utilizes a set of custom-made, solid, homogenous phantoms (Fig. 1a), with gradually increasing absorption and same constant scattering (with the main purpose of also testing linearity response of any device, as mentioned). The in-house recipe was developed and refined over the years at UCL[31,32] to manufacture solid, homogenous phantoms in large numbers, rapidly and relatively inexpensively (compared to commercial ones), also allowing for fine tuning of their absorption and scattering properties. The recipe consists of using polyester resin as the main bulk matrix, a NIR dye as the absorption agent, and a polyester pigment as the scattering agent. These components are mixed electrically in appropriate concentrations and then casted in a polypropylene (PP) plastic mould with shape and size desired. After addition of a hardener catalyst (1-2% in weight), the mixture of resin, absorbing and scattering agents will solidify in the desired homogeneous phantom, typically in a few hours to half a day of rest at room temperature, under a chemical fumes hood. The ingredients of the recipe are reported in Table 3.

The total cost for making a batch of solid phantom mixture (1 kg) with fixed optical properties is below 20£, with each batch capable of casting approximately a single, brick-shaped phantom of volume of approximately 945 $cm^3$ (The density of the resin being ~1.06 $g/cm^3$).

**Table 3** Components of the recipe for manufacturing solid, homogeneous phantoms used in PROTOCOL 1.

| Component | Brand and model | Role | Quantity for one single phantom | Average cost |
|---|---|---|---|---|
| Polyester resin | Tiranti, Clear Casting Resin AM | Bulk matrix | 1 kg | £17.49 per 1 kg |
| Hardener | Tiranti, Liquid Hardener | Hardener for the resin | 1-1.5 mL | £5.99 per 50 mL |
| NIR dye | Pro jet, 900NP | Absorbing agent | 0.1 – 0.3 g | £2 per 1 g |
| Polyester pigment | Tiranti, Super White | Scattering agent | 1 g | £5.83 for 100 g |

NIR: Near-infrared.

For the scope of PROTOCOL 1, three rectangular solid, homogenous phantoms of size 19 x 13 x 4 cm$^3$ were produced (Solid phantom #1, #2 and #3), all with the same scattering (1 g of polyester pigment per kg of resin is estimated to produce a reduce scattering coefficient of approximately 7.5 cm$^{-1}$ at 800 nm), and with linearly increasing absorption, with theoretical absorption coefficients of 0.1, 0.25 and 0.8 cm$^{-1}$ at 800 nm, respectively (these were obtained by diluting 0.195, 0.4875 and 1.56 g of NIR dye per kg of resin).

Each phantom was then characterised using MAESTROS II[4], a multi-wavelength, TD-NIR system developed at UCL, to measure their experimental optical properties at the reference wavelength of 800 nm (Fig. 1a). These experimental optical properties are reported in Table 4, against the theoretical ones from the recipe.

**Table 4** Comparison between the theoretical and the measured optical properties of the three manufactured phantoms used in PROTOCOL 1.

| Solid Phantom # | Reference wavelength | Theoretical absorption, $\mu_a$ | Theoretical scattering, $\mu'_s$ | Measured absorption, $\mu_a$ | Measured scattering, $\mu'_s$ |
|---|---|---|---|---|---|
| 1 | 800 nm | 0.1 cm$^{-1}$ | 7.5 cm$^{-1}$ | 0.11 cm$^{-1}$ | 6.74 cm$^{-1}$ |
| 2 | 800 nm | 0.25 cm$^{-1}$ | 7.5 cm$^{-1}$ | 0.26 cm$^{-1}$ | 6.94 cm$^{-1}$ |
| 3 | 800 nm | 0.8 cm$^{-1}$ | 7.5 cm$^{-1}$ | 0.81 cm$^{-1}$ | 6.44 cm$^{-1}$ |

The characterisation of the three manufactured phantoms using MAESTROS II showed that the measured absorption and scattering properties of the custom-made solid phantoms are within a within a 10% and 15% of difference error, respectively, compared to the theoretical values expected from the recipe. This demonstrates the high reproducibility and accuracy of the recipe for PROTOCOL 1.

*2.3 PROTOCOL 2: Physiological monitoring assessment with homogeneous liquid phantom*

PROTOCOL 2 is based on a physiologically realistic recipe for a homogeneous, dynamic liquid phantom aimed at recreating, in a controllable manner, the exact same type of optical contrast occurring in the placenta and adjacent tissue from dynamic changes in haemodynamics, oxygenation and metabolism over time. Such changes are connected to the real optical signatures of haemoglobin (both $HbO_2$ and $HHb$) and the oxidised state of CCO (oxCCO), which act as biomarkers of these physiological processes.

The homogenous, liquid phantom recipe for PROTOCOL 2 uses an established formula developed at UCL[33,34], composed of (1) a phosphate-buffered saline (PBS) and water matrix, Intralipid 20% (for scattering and lipid absorption), human or animal blood (to induce the oxygenation contrast), and yeast (for the metabolic contrast). The mixture (Fig. 1c) is housed in a custom-made Plexiglas container, of dimensions $15 \times 15 \times 15$ cm$^3$, with 11 x 8 cm windows of Mylar foil (100 μm in thickness, with negligible refraction and absorption in the NIR range) on two opposite lateral sides, that are used to transmit the signal to and from any optical probe without having the instrumentation to be in contact with the liquid, making the measurement procedures in PROTOCOL 2 completely waterproof,  as well as allowing for multiple probes to measure the same mixture simultaneously. Dissolved oxygen ($DO_2$) percentage within the solution is monitored constantly using a calibrated oxygen probe. Solution pH is concurrently monitored using a pH probe and maintained at about physiological values (7.2-7.4). The entire setup is placed on a hot stirring plate equipped with a temperature feedback system, that is set to maintain a constant 36-37°C. Constant stirring at about 520 RPM ensures that the solution remains homogeneous and prevent the formation of clusters (flocking) of yeast and Intralipid that may change the scattering properties of the mixture over time.

The basic principle of working of the liquid phantom for PROTOCOL 2 is that the oxygenation states of the blood can be fully controlled and set to 100% by bubbling $O_2$ in the solution. This will result in the haemoglobin to be fully oxygenated (increase in $HbO_2$, decrease in HHb). Then, the user can induce a deoxygenation of the solution by bubbling $N_2$ in the solution (decrease in $HbO_2$, increase in HHb) Thus, by alternatively bubbling $O_2$ and $N_2$ in the solution, one can control the oxygenation contrast in the solution. Otherwise, when adding yeast, no $N_2$ is needed, as the yeast will consume the oxygen in the solution and reduce the oxygenation of the blood. Additionally, as the oxygen level drops, the oxidative state of CCO (oxCCO) in the yeast will also change, providing the active metabolic contrast (increase of oxCCO during oxygenation, decrease during deoxygenation). The composition are quantities of the homogenous, liquid phantom recipe for PROTOCOL 2 are reported in Table 5, for both the cases when CCO is present (by addition of yeast as deoxygenating agent) and when it is not (deoxygenation by $N_2$ bubbling).

The purpose of testing any photonics instrumentation with PROTOCOL 2 is to quantitatively and systematically assess its performances in reconstructing dynamic changes in oxygenation and metabolism over time, akin to what would occur in the placenta *in vivo*, by introducing the exact same biomarkers and physiological variations that are naturally occurring in human tissues. Furthermore, by evaluating the results of the testing on both the cases when the deoxygenation is induced with and without yeast, it is also possible to assess and validate the absence of crosstalk between the metabolic and oxygenation contrast (the phantom deoxygenated with $N_2$ acting as an invariant against the changes in oxCCO).

**Table 5** Compositions and main parameters of both iterations of the liquid optical phantom used in PROTOCOL 2.

| Composition | Phantom with yeast | Phantom without yeast |
|---|---|---|
| Deionised water | 1.4 L | 1.4 L |
| Intralipid 20% | 75 g | 75 g |
| PBS | 50 mM | 50 mM |
| Blood | 25 mL | 25 mL |
| Yeast | 5 g | 0 |
| Oxygenation | $O_2$ bubbling | $O_2$ bubbling |
| Deoxygenation | Yeast | $N_2$ bubbling |

PBS: Phosphate-buffered saline.

*2.4 PROTOCOL 3: Depth sensitivity assessment with multilayered, hybrid phantom*

PROTOCOL 3 is based on a multilayered, hybrid (both solid and liquid) phantom intended at realistically modelling the layered structure of the abdominal wall and the placenta, as well as their corresponding optical properties, as depicted in Fig. 1d. The basal layer of this hybrid phantom is composed of the same recipe of homogenous, dynamic liquid phantom as used in PROTOCOL 2. This simulates the placenta itself, by controlling also its oxygenation and metabolic states in the same manners as described in Sec. 2.3. On the top of the placenta-proxy, dynamic liquid phantom, a two-layers solid phantom based on the same recipe as used in PROTOCOL 1 is placed above the mylar window of the housing (Fig. 1e) to replicate any desired thicknesses and the optical properties of the overlayers above the placenta, namely (i) skin and adipose tissue, and (ii) the muscle (uterus/rectus).

The optical properties of the two layers of the solid phantom remain fixed, whilst only the liquid phantom at the bottom will provide changes over time, in the same fashion as in PROTOCOL 2. The two solid overlayers of PROTOCOL 3 are produced by casting two batches of solid, homogeneous phantoms using the same procedures described per PROTOCOL 1 in Sec. 2.2, each with the required optical properties for simulating realistically a combination of skin and adipose tissue (Layer #1) and muscle tissue (Layer #2), respectively. Each batch is then cut along the longest size to produce the two solid overlayers of any desired thickness, to test different levels of depth sensitivity across the range of the average anatomical distances from top of the surface of the skin to the top of the placenta (using ultrasound scans as guidance). The optical properties of both the solid overlayers used in PROTOCOL 3 are reported in Table 6, with examples of thicknesses for each layer relative to a total placental depth within the abdomen of 1.5 cm. These are compared against average theoretical estimates of the optical properties of the different types of tissue as available in literature[29], as reported in Table 7.

Table 6 Example of experimental optical properties and relative thicknesses of the solid overlayers used in PROTOCOL3 for the chosen case of placental depth of 1.5 cm.

| Solid Layer # | Thickness | Reference wavelength | Absorption, $\mu_a$ | Reduced scattering, $\mu'_s$ |
|---|---|---|---|---|
| 1 | 0.8 cm | 800 nm | 0.05 cm$^{-1}$ | 11 cm$^{-1}$ |
| 2 | 0.7 cm | 800 nm | 0.11 cm$^{-1}$ | 8 cm$^{-1}$ |

Table 7 Average optical properties of the overlayers of the abdominal wall above the placenta[29].

| Overlayer | Reference wavelength | Absorption, $\mu_a$ | Reduced scattering, $\mu'_s$ |
|---|---|---|---|
| Skin | 800 nm | 0.055 cm$^{-1}$ | 13 cm$^{-1}$ |
| Adipose | 800 nm | 0.003 cm$^{-1}$ | 13 cm$^{-1}$ |
| Muscle | 800 nm | 0.1 cm$^{-1}$ | 8 cm$^{-1}$ |

The latter thickness of 1.5 cm has been taken as an example here since it represents the median value of placental depth across ultrasound scans performed on 268 healthy participants during maternal gestation at the Institute for Women's Health in the University College London Hospital (UK), at different gestational stages (ranging from 24 to 40 weeks)[27]. The measurements were approved by the local ethics committee of University College London Hospital and the participating patients signed written consent.

The purpose of PROTOCOL 3 is to test the depth sensitivity of any optical system for placental monitoring, as well as to provide a more realistic scenario aimed at simulating the configuration of the *in vivo* measurements on the placenta, albeit in a controllable and validated manner. The light from the device will have to penetrate the attenuation provided by the solid overlayers and be able to sense the changes in physiology occurring on the bottom liquid phantom. Then, enough signal will have to bounce back to the sensor for the physiological signatures to be detected and accurately reconstructed. Thus, PROTOCOL 3 has the capability to quantitatively assess such performances, ultimately providing the most effective scenario to validate metrologically any photonics device before being applied in the clinics.

## 3  Examples of application of OptoCENTAL

Four NIRS instruments developed at UCL for placental monitoring have been enrolled and employed to provide examples of application of the OptoCENTAL platform, as described here:

- The microCYRIL[35] is a wearable, broadband, CW-NIRS device (Fig. 2a), based on broadband LED illumination (400-1000 nm). It utilizes two sources (LED A and B) linearly spaced at 3 cm and 6 cm SDS from a miniaturised slit-spectrometer, operating in the range between 640 and 1050 nm, for full spectra collection.

- The FetalSenseM V1.5[36] is a wearable, multi-wavelength, CW-NIRS device (Fig. 2b), based on LED illumination at six NIR wavelengths (780, 810, 830, 840, 850 and 890 nm). It utilizes two sources (LED A and B) centrally located on the sensor body at 2 cm from each other, and two photodiodes (PD1 and PD2) placed symmetrically from the sources, providing two sets of two source-detector separations (SDS) at 3 and 5 cm.

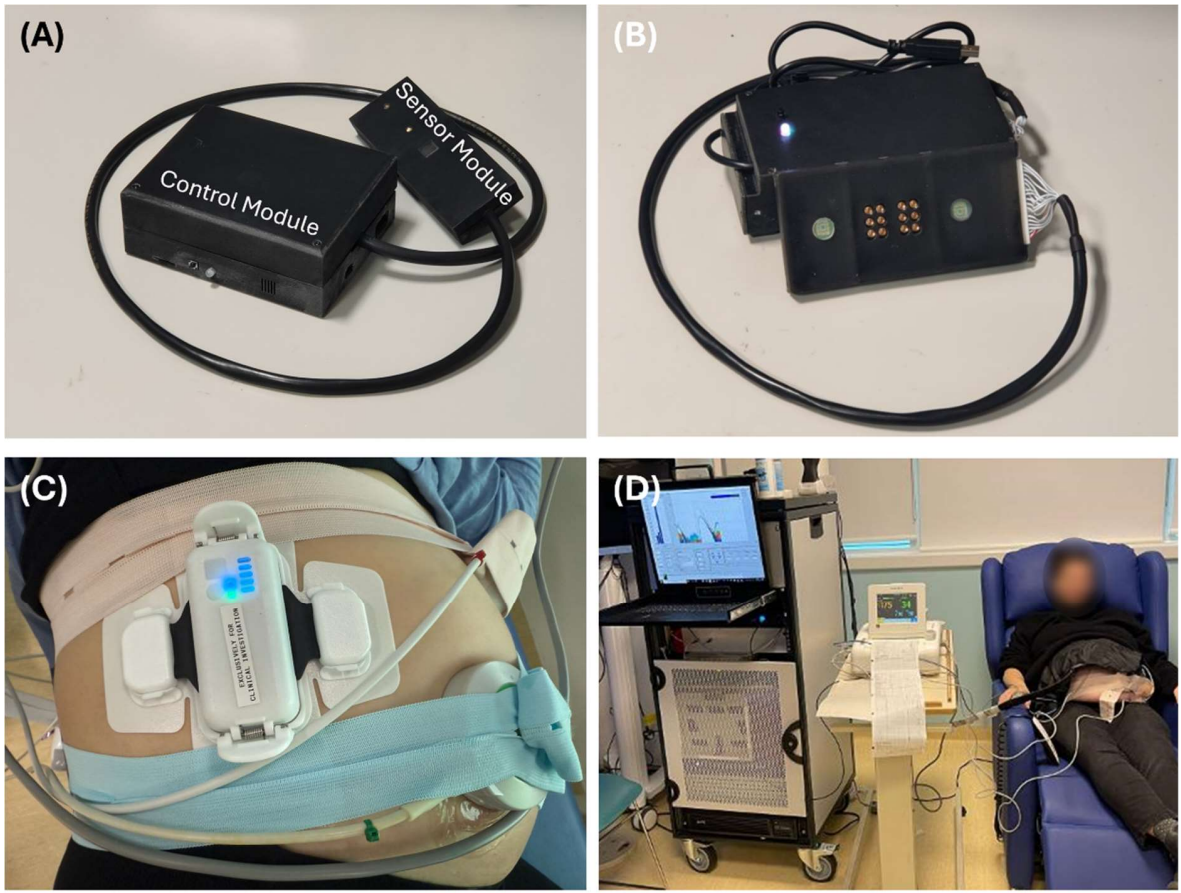

**Fig. 2** Pictures of the four placental-monitoring, optical instruments developed ta UCL and used as exemplary application of the OptoCENTAL platform: (A) microCYRIL is a wearable, broadband, CW-NIRS device; (B) FetalSenseM V1.5 is a wearable, six-wavelengths, CW-NIRS device; (C) FetalSenseM V2 is a wearable and wireless, six-wavelengths, CW-NIRS device; (D) MAESTROS II is a transportable, 16-wavelenghts, TD-NIRS system.

- The FetalSenseM V2 is a wireless and wearable, multi-wavelength, CW-NIRS device (Fig. 2c), based on LED illumination at six NIR wavelengths (782, 800, 818, 830, 848 and 888 nm). It utilizes two sources (LED A and B) located on the flexible wings of the sensor body, and three photodiodes (PD1, PD2 and PD3) placed symmetrically in its centre at 1 cm from each other. The configuration provides two sets of three SDS at 3, 4 and 5 cm.

- The MAESTROS II[4] is a transportable, multi-wavelength, TD-NIRS system (Fig. 2d), based on pulsed, supercontinuum laser (SCL) illumination, filtered via acousto-optics tuneable filters (AOTF) and delivered via a single optical fibre. This allows the user to select up to 16 NIR wavelengths from 780 to 870 nm, at 6-nm steps. Detection is performed via four optical fibres, directing the reflected light from the tissues into hybrid photomultiplier tubes (PMT) and a time-correlated single photon counting (TCSPC) unit. Four SDS can thus be employed at 2, 3, 4 and 5 cm.

All the four instruments allow users to estimate dynamic, relative changes in concentration of $HbO_2$, HHb and oxCCO via the UCLn algorithm, using a wavelength-dependent, differential pathlength factor (DPF)[13,37,38]. MAESTROS II can also infer absolute changes in the three chromophores via temporal point spread function (TPSF) fitting[4,33].

Table 8 Main features and technical specification of the four NIRS instruments for placental monitoring employed to provide exemplary application of the OptoCENTAL platform.

| Instrument | Type | Wavelengths | Spectral resolution (FWHM) | Number of sources | Number of detectors | SDS | Algorithms |
|---|---|---|---|---|---|---|---|
| microCYRIL | Broadband, wearable, CW-NIRS | Broadband (640-1050) | 12 nm | 2 | 1 | 3, 6 cm | UCLn, SRS |
| FetalSenseM V1.5 | Multi-wavelength, wearable, CW-NIRS | 6 (780, 810, 830, 840, 850 and 890 nm) | 25-44 nm | 2 | 2 | 3, 5 cm | UCLn, SRS, DS |
| FetalSenseM V2 | Multi-wavelength, wearable and wireless, CW-NIRS | 6 (782, 800, 818, 830, 848 and 888 nm) | 18-30 nm | 2 | 3 | 3, 4, 5 cm | UCLn, SRS, DS |
| MAESTROS II | Multi-wavelength, TD-NIRS | 16 (780-870 nm at 6-nm steps) | 5-7 nm | 1 | 4 | 2, 3, 4, 5 cm | UCLn, TPSF fitting |

SDS: Source-detector distance; CW: Continuous-wave; TD: Time-domain; NIRS: Near-infrared spectroscopy; FWHM: Full-width, half maximum; SRS: Spatially-resolved spectroscopy; DS: Dual Slope; TPSF: Temporal point spread function.

The dual set of SDS provided by microCYRIL, FetalSenseM V1.5 and V2 also enables the use of multi-distance algorithms, such as spatially-resolved spectroscopy (SRS)[39], for all three CW-NIRS systems, and Dual Slope (DS)[40], only for FetalSenseM V1.5 and V2 (thanks to their dual-source configurations), all for the purpose of measuring absolute tissue oxygen saturation (StO$_2$), whereas MAESTROS II can recover StO$_2$ again via TPSF fitting. The main technical features of each of the four NIRS instruments are summarised in Table 8.

*3.1 Examples of application of PROTOCOL 0*

For the FetalSenseM V1.5, detectors in the simulated model from PROTOCOL 0 were placed 3 and 5 cm from each source in a symmetrical design, whereas for the FetalSenseM V2, the detectors were placed 3, 4 and 5 cm from each source in a symmetrical design. The MAESTROS II detectors were positioned 2, 3, 4 and 5 cm from the source. Each detector was modelled as a 1-mm disk positioned on the surface of the volume.

The primary goal of PROTOCOL 0 is to computationally assess the sensitivity of the proposed optical configuration to scan the placenta layers. Thus, to compute the sensitivity profile of each configuration and volume the adjoint method was used[41]. This differed slightly between the two forms of NIRS device. For the CW-NIRS cases (FetalSenseM V1.5 and V2), the procedure begins by computing the fluence rate using MCX a given source to each detector. Then the source position is moved to each detector position and the fluence is re-computed. For each detector position, these two outputs are multiplied by each other. This provides the final sensitivity profile for a given source and detector. Once obtained these were then normalised by the sum of the sensitivity profile to obtain the probability volume of photon propagation between the source and a given detector. To then obtain the sensitivity of the configuration to each tissue layer, the sensitivity probability was summed within each layer and then divided by the total sensitivity probability. For the TD-NIRS case (MAESTROS II) the procedure differed firstly, in that the sensitivity profile is first multiplied by the time-step used in order to assess the sensitivity of different time-windows of 0.02 ns each in duration. Then, after re-positioning the source at each detector, the original sensitivity profile is convolved, rather than multiplied, by the sensitivity profile at each detector. The final sensitivity for each tissue layer at each time-window is then computed in the same way as the CW-NIRS case.

Example sensitivity profile for each of the devices tested are reported in Figure 3 and Figure 4. Figure 3 is an example sensitivity assessment for the MAESTROS II system. Each subplot provides the sensitivity associated with a given detector, at each tissue layer and time-window. The associated numerical values can be found in the supplementary materials.

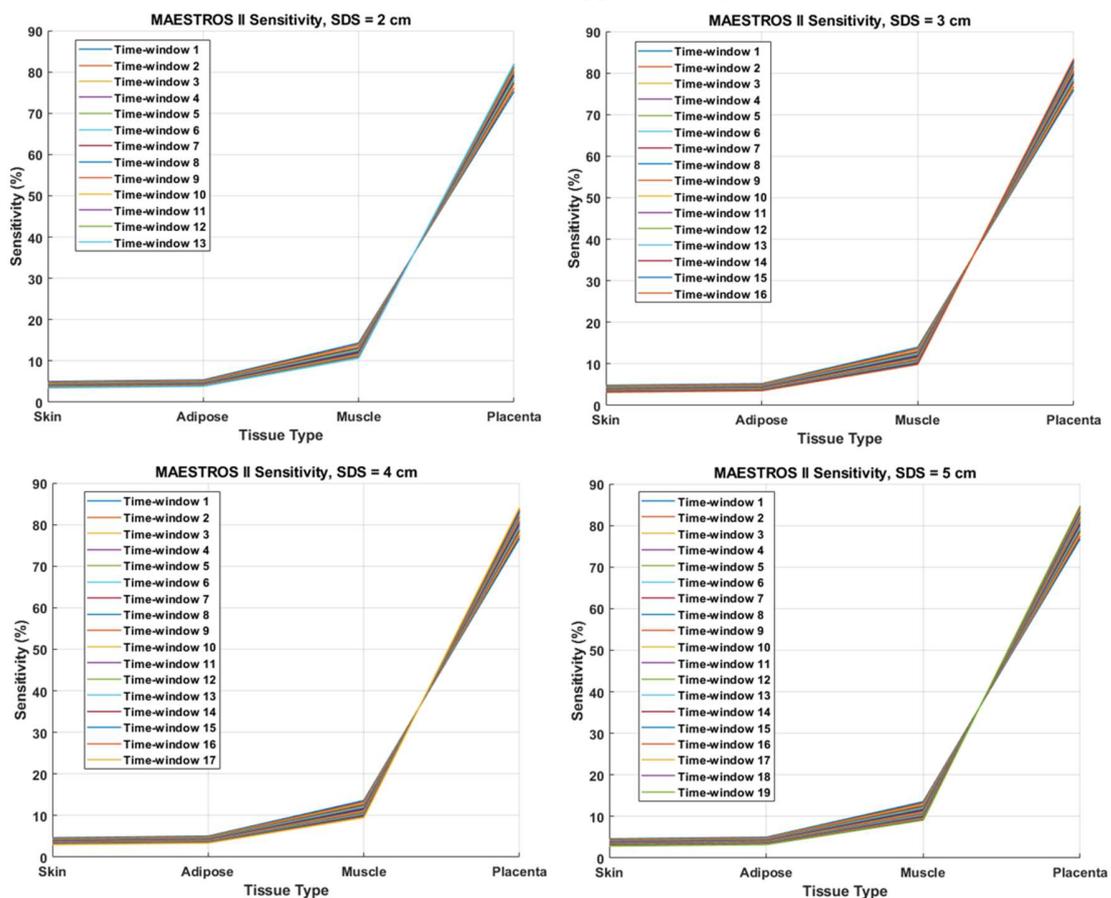

**Fig. 3**. Example sensitivity profile for MAESTROS II, for all its four SDS channels (2, 3 4, and 5 cm), and for each 0.02-ns time windows in the TD acquisition. The corresponding data table is shown in the supplementary materials.

Figure 4 is an example of assessment for the both the FetalSenseM V1.5 and V2 devices, across all their corresponding SDS channels. The associated numerical values can be found in the supplementary materials.

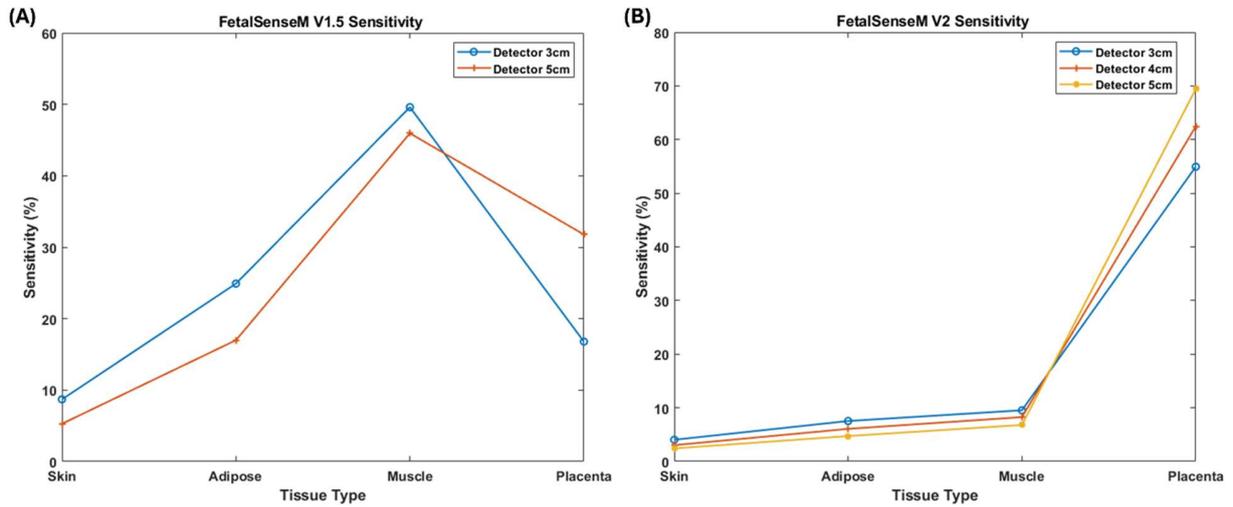

**Fig. 4**. (A) Example sensitivity profile for FetalSenseM V1.5, for a both SDS channels, at 3 and 5 cm; (B) Example sensitivity profile for FetalSenseM V2, for all its three SDS channels, at 3, 4 and 5 cm. The corresponding data tables are shown in the supplementary materials.

*3.2 Examples of application of PROTOCOL 1*

PROTOCOL 1 is composed of five assessment procedures of basic instrument performances: (1) SNR; (2) noise/variance; (3) linearity; (4) stability; and (5) reliability/repeatability.

SNR is a measure that compare the level of signal detected by any tested optical system to the level of instrumental and background noise (the intensity detected by any sensor of the systems when the main illumination sources are off). For PROTOCOL 1, SNR is defined by the ratio between the average light intensity detected from any of the custom-made, homogenous, solid phantom (described in Sec. 2.2) in each SDS channel of the sensors over the average intensity detected when the instrument illumination is absent. The highest the value of the SNR, the better are the performances of the devices when applied to tissue-equivalent phantoms.

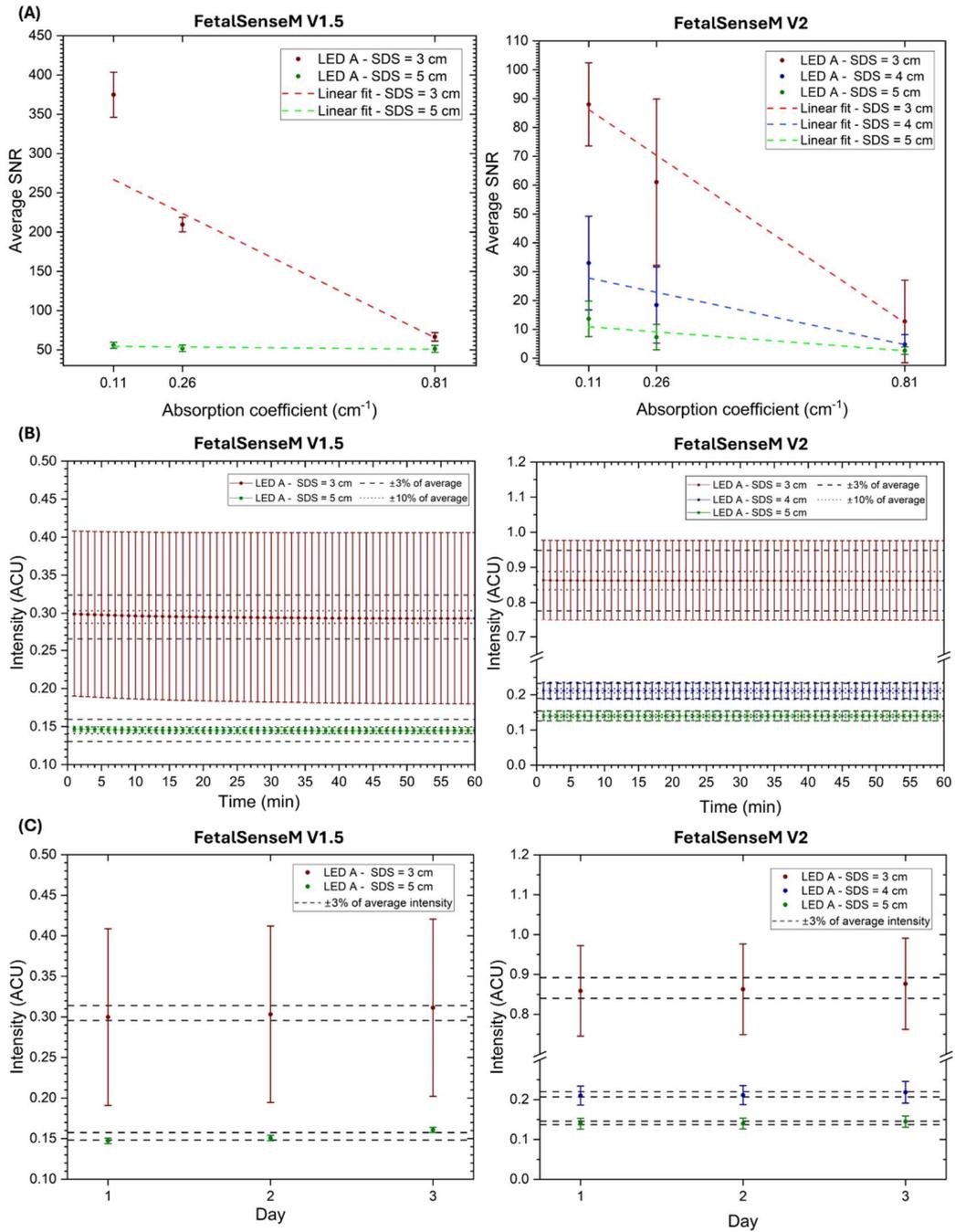

**Fig. 5** Examples of results of the application of PROTOCOL 1 on the FetalSenseM V.15 and v2 devices for placental monitoring: (A) Linearity profiles of the average SNR across the three solid phantoms (Phantom #1 to #3); (B) Stability profiles of the average intensity on Phantom #1 across repeated measurements over 1 hour, where dotted and dashed line shows ±3% and ±10% changes with respect to the average value; (C) Repeatability assessment across three consecutive days of measurements on the same phantom (Phantom #1) and with the same experimental settings, where dashed line shows ±3% changes with respect to the average value.

As an example, measurements of SNR during PROTOCOL 1 were conducted with both FetalSenseM V1.5 and V2 on all the available solid phantoms (Solid phantom #1 to #3, as per Table 4). The results of the SNR assessments are reported in Table 9.

Table 9 Comparison of SNR assessment for FetalSenseM V1.5 and V2 across different SDS on all the three solid phantoms (Phantom #1 to #3) during PROTOCOL 1.

| Phantom # | Average SNR ± standard deviation - FetalSenseM V1.5 | | | |
|---|---|---|---|---|
| | LED A | | LED B | |
| | SDS = 3 cm | SDS = 5 cm | SDS = 3 cm | SDS = 5 cm |
| 1 | 372.194 ± 147.72 | 74.349 ± 16.76 | 365.871 ± 160.62 | 77.080 ± 24.95 |
| 2 | 236.930 ± 51.80 | 60.176 ± 7.20 | 242.588 ± 60.34 | 67.580 ± 16.04 |
| 3 | 74.106 ± 7.11 | 52.589 ± 0.85 | 77.105 ± 8.55 | 51.726 ± 2.76 |

| Phantom # | Average SNR ± standard deviation - FetalSenseM V2 | | | | | |
|---|---|---|---|---|---|---|
| | LED A | | | LED B | | |
| | SDS = 3 cm | SDS = 4 cm | SDS = 5 cm | SDS = 3 cm | SDS = 4 cm | SDS = 5 cm |
| 1 | 99.547±40.09 | 25.454±11.78 | 12.305±5.09 | 116.015±44.47 | 26.797±11.83 | 13.721±5.38 |
| 2 | 50.212±25.90 | 13.214±8.50 | 6.165±2.97 | 43.572±24.34 | 11.964±8.68 | 5.836±3.17 |
| 3 | 10.850±10.65 | 3.702±2.06 | 2.387±0.94 | 10.666±10.34 | 3.354±2.03 | 2.386±1.00 |

SDS: Source-detector distance.

The noise/variance assessment in PROTOCOL 1 involves the quantification of the variability in the readings of any tested instrument on the set of solid phantoms, due to random effects. Such assessment has the purpose of evaluating the susceptibility of the systems to statistical fluctuations in their readings. Noise/variance can be quantified by looking at the coefficient of variation (CV) in the readings over a determined acquisition time, and its defined as in Eq. 6:

$$\text{CV} = \frac{\sigma(x)}{\langle x \rangle} \tag{6}$$

where $\sigma(x)$ is the standard deviation in the light intensity reading $x$ during the time of acquisition,

and $\langle x \rangle$ denotes its average value. The lowest is the value of CV, the higher is the sensitivity of any placental-monitoring instrument in capturing small changes in the optical properties of the tissues they investigate.

Measurement of CV during PROTOCOL 1 were again conducted with both FetalSenseM V1.5 and V2 on all the solid phantoms (Phantom #1 to #3). The results are reported in Table 10.

Table 10 Comparison of noise assessment estimating CV values for FetalSenseM V1.5 and V2 across different SDS on all the three solid phantoms (Phantom #1 to #3) during PROTOCOL 1.

| Phantom # | Coefficient of variation (CV) - FetalSenseM V1.5 | | | |
|---|---|---|---|---|
| | LED A | | LED B | |
| | SDS = 3 cm | SDS = 5 cm | SDS = 3 cm | SDS = 5 cm |
| 1 | $1.087 \cdot 10^{-2}$ $\pm 1.29 \cdot 10^{-2}$ | $4.977 \cdot 10^{-3}$ $\pm 4.12 \cdot 10^{-3}$ | $1.478 \cdot 10^{-3}$ $\pm 5.42 \cdot 10^{-4}$ | $2.267 \cdot 10^{-3}$ $\pm 9.32 \cdot 10^{-4}$ |
| 2 | $2.516 \cdot 10^{-3}$ $\pm 6.88 \cdot 10^{-4}$ | $1.776 \cdot 10^{-3}$ $\pm 1.49 \cdot 10^{-4}$ | $1.568 \cdot 10^{-3}$ $\pm 5.45 \cdot 10^{-4}$ | $1.850 \cdot 10^{-3}$ $\pm 2.21 \cdot 10^{-4}$ |
| 3 | $1.643 \cdot 10^{-3}$ $\pm 3.62 \cdot 10^{-4}$ | $1.086 \cdot 10^{-3}$ $\pm 8.43 \cdot 10^{-4}$ | $2.153 \cdot 10^{-3}$ $\pm 6.34 \cdot 10^{-4}$ | $1.520 \cdot 10^{-3}$ $\pm 3.99 \cdot 10^{-4}$ |

| Phantom # | Coefficient of variation (CV) - FetalSenseM V2 | | | | | |
|---|---|---|---|---|---|---|
| | LED A | | | LED B | | |
| | SDS = 3 cm | SDS = 4 cm | SDS = 5 cm | SDS = 3 cm | SDS = 4 cm | SDS = 5 cm |
| 1 | $4.843 \cdot 10^{-4}$ $\pm 1.33 \cdot 10^{-4}$ | $8.107 \cdot 10^{-4}$ $\pm 1.88 \cdot 10^{-4}$ | $1.357 \cdot 10^{-3}$ $\pm 2.27 \cdot 10^{-4}$ | $9.697 \cdot 10^{-4}$ $\pm 1.11 \cdot 10^{-4}$ | $1.365 \cdot 10^{-3}$ $\pm 3.24 \cdot 10^{-4}$ | $1.123 \cdot 10^{-3}$ $\pm 2.25 \cdot 10^{-4}$ |
| 2 | $2.240 \cdot 10^{-3}$ $\pm 1.32 \cdot 10^{-3}$ | $4.946 \cdot 10^{-3}$ $\pm 2.09 \cdot 10^{-3}$ | $6.067 \cdot 10^{-3}$ $\pm 2.24 \cdot 10^{-3}$ | $1.697 \cdot 10^{-3}$ $\pm 1.07 \cdot 10^{-3}$ | $3.551 \cdot 10^{-3}$ $\pm 1.47 \cdot 10^{-3}$ | $4.425 \cdot 10^{-3}$ $\pm 1.52 \cdot 10^{-3}$ |
| 3 | $9.340 \cdot 10^{-3}$ $\pm 3.98 \cdot 10^{-3}$ | $1.181 \cdot 10^{-2}$ $\pm 4.35 \cdot 10^{-3}$ | $2.246 \cdot 10^{-2}$ $\pm 7.62 \cdot 10^{-3}$ | $9.025 \cdot 10^{-3}$ $\pm 6.06 \cdot 10^{-3}$ | $1.112 \cdot 10^{-2}$ $\pm 5.72 \cdot 10^{-3}$ | $1.372 \cdot 10^{-2}$ $\pm 5.23 \cdot 10^{-3}$ |

SDS: Source-detector distance.

Linearity in the measurements conducted in PROTOCOL 1 can be assessed by comparing the outcomes of the readings of any optical system from the set of three phantoms with varying optical properties. Linearity plots can then be obtained for SNR by graphing their average values at the reference wavelengths (800 nm) obtained with the sets of phantoms with different properties, as a function of their corresponding absorption coefficients (Table 4). Examples of these linearity plots are depicted in Figure 5a for both FetalSenseM V1.5 and V2. Pearson correlation coefficients R have been calculated from the linearity profile to quantify linearity as a correlation between the average SNR and the absorption coefficients of the three solid phantoms (Solid phantom #1 to #3), with perfect linearity corresponding to R = -1. The coefficients $R$ are then reported in Table 11 for the linearity profiles of both FetalSenseM V1.5 and V2.

Table 11 Examples of assessment of linearity with Pearson correlation coefficients R (where perfect linearity is given by R = -1) for FetalSenseM V1.5 and V2 during PROTOCOL 1.

| Correlation coefficients $R$ of linearity profiles - FetalSenseM V1.5 ||||
|---|---|---|---|
| LED A || LED B ||
| SDS = 3 cm | SDS = 5 cm | SDS = 3 cm | SDS = 5 cm |
| -0.7719 | -0.8395 | -0.8143 | -0.8773 |

| Correlation coefficients $R$ of linearity profiles - FetalSenseM V2 ||||||
|---|---|---|---|---|---|
| LED A ||| LED B |||
| SDS = 3 cm | SDS = 4 cm | SDS = 5 cm | SDS = 3 cm | SDS = 4 cm | SDS = 5 cm |
| -0.9878 | -0.9442 | -0.9189 | -0.8716 | -0.8942 | -0.9710 |

SDS: Source-detector distance.

Stability assessment in PROTOCOL 1 involves the testing of the capability of any optical instrument of providing consistent reading over multiple repetitions in short time durations, i.e., by repeating multiple measurements at subsequent time intervals on the same solid phantom and without varying the experimental conditions. The stability assessment has the purpose of revealing any potential short- or long-term drifts in the devices, as well as any unwanted fluctuations over time which may give rise to false positives or negatives in the physiological monitoring of the placenta. In PROTOCOL 1 of OptoCENTAL, stability is assessed over a duration of 1 hour of repeated measurements on Solid phantom #1 at intervals of 1 min. Variability in stability is then evaluated against threshold values of ±3% and ±10% of the average intensity over the whole hour of measurements. Examples of results of the stability assessment in PROTOCOL 1 are reported in Figure 5b, as stability profiles of the readings over time.

Finally, reliability (or repeatability) in PROTOCOL 1 measures the level of reproducibility and consistency of the readings of any tested instrument across measurements of the same target phantom over three different days. It can be expressed as the percentage difference between average measurements across three consecutive days compared to their mean value. The reliability/repeatability assessment have the purpose of quantifying how placental-monitoring systems are self-consistent and allow users for correlation of results obtained in different measurements sessions. As an example, for both FetalSenseM V1.5 and V2, repeated measurements were carried out on Solid phantom #1 at intervals of 1 day for a total duration of 3 consecutive days at the same experimental conditions and setting. Variability in repeatability is then evaluated against a threshold value of ±3% of the average intensity over the entire three days of measurements. The results are reported in Figure 5c.

*3.3 Examples of application of PROTOCOL 2*

A first example of application of PROTOCOL 2 was performed on microCYRIL, as bNIRS is the typically "gold standard" approach to minimise crosstalk in the simultaneous estimates of $HbO_2$, HHb and CCO[13,42]. In the first cycle of PROTOCOL 2, firstly a baseline on the liquid phantom was established without yeast at full oxygenation state ($DO_2 \geq 100\%$), followed by a full deoxygenation ($DO_2 \sim 0\%$) induced via $N_2$ bubbling. Afterwards, return to full re-oxygenation ($DO_2 \geq 100\%$) was achieved via bubbling of $O_2$. Following this first cycle of deoxygenation and re-oxy-genation, a second cycle was conducted with yeast added to the mixture to induce again full deoxygenation ($DO_2 \sim 0\%$), and subsequently full re-oxygenation ($DO_2 \geq 100\%$) was re-established via bubbling of $O_2$. Dynamic, relative changes $\Delta[HBO_2]$, $\Delta[HHb]$ and $\Delta[oxCCO]$ in the concentrations of $HbO_2$, HHb and oxCCO were calculated for each dataset acquired with microCYRIL using the UCLn algorithm. The results of the data analysis on PROTOCOL 2 for microCYRIL are reported in Figure 6, for the first cycle without yeast (Figure 6a) and the second cycle with the addition of yeast (Figure 6b).

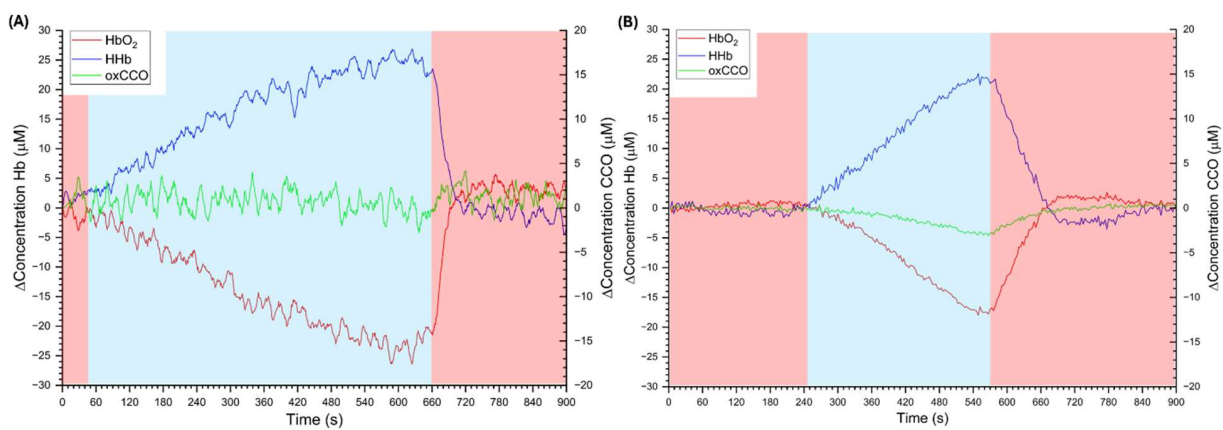

**Fig. 6** Dynamic changes in concentration of $HbO_2$ (red), HHb (blue) and oxCCO (green) over time during PROTOCOL 2, for microCYRIL at SDS of 3 cm. Two cycles are shown: (A) one where deoxygenation is induced via $N_2$ bubbling and no yeast is present; (B) the other where deoxygenation is achieved via yeast. The comparison of the two allows to validate the absence of crosstalk from CCO in the measurements with microCYRIL.

The dynamic changes in concentrations of all the chromophores of interest retrieved by microCYRIL show consistent trends in accordance with the expected physiological changes in the liquid phantom during PROTOCOL 2 for both cycles. The concentration of $HbO_2$ decreases during deoxygenation events (max at 22.60 μM) and increases during reoxygenation, while the opposite trend is reported for the concentration of HHb (max at -18.00 μM). The concentration of oxCCO is also reported to decrease during the deoxygenation via yeast (max at -2.85 μM) in the second cycle (Figure 6b), as expected. As expected, PROTOCOL 2 also successfully demonstrates that negligible crosstalk errors (max crosstalk error equal to 0.05 µM) are quantified for microCYRIL during the first cycle of deoxygenation when yeast is not present (Figure 6a).

A second example of application of PROTOCOL 2 was performed with both FetalSenseM V2 and MAESTROS II simultaneously, by virtue of the multiple access windows on the sides of the liquid phantom housing. Again, dynamic, relative changes $\Delta[HBO_2]$, $\Delta[HHb]$ and $\Delta[oxCCO]$ in the concentrations of $HbO_2$, HHb and oxCCO were calculated for each dataset acquired with FetalSenseM V2 and MAESTROS II using the UCLn algorithm for both instruments. The same wavelength-dependent DPF values at time 0 were used for both datasets, although MAESTROS II implemented also a time-varying DPF calculated from TPSF fitting for each time point of the acquisition. This produces scaling differences in the magnitude of the responses of $HbO_2$, HHb and oxCCO in the comparison between the two systems The results of the data analysis on PROTOCOL 2 are reported in Figure 7 for the same, comparable SDS of 3 cm for both systems.

The dynamic changes in concentrations of all the chromophores of interest again show consistent trends in accordance with the expected physiological changes in the liquid phantom during PROTOCOL 2, for both tested optical devices. However, this time PROTOCOL 2 successfully identifies and quantify crosstalk errors (mean crosstalk error equal to -1.66 µM, max crosstalk error equal to -3.32 µM) for the FetalSenseM V2 during the first phase of deoxygenation when yeast is not present. This is contrarily to MAESTROS II, which exhibits no significant crosstalk in the same phase (mean crosstalk error equal to 0.31 µM).

Reconstruction of absolute oxygen saturation $StO_2$ during PROTOCOL 2 was also implemented, using both the SRS and dual slope algorithms for FetalSenseM V2, as well as via broadband TPSF fitting for MAESTROS II. A comparison of the results of the absolute oximetry quantification (Fig. 7c) for both optical system during PROTOCOL 2 showed that the greatest accuracy is achieved with MAESTROS II, as in accordance with literature regarding the efficiency and performances of TD-NIRs compared to CW-NIRS[43]: Indeed, the TD-NIRs system is capable of recovering full oxygenation up to 98.34% and deoxygenation down to about 27.12% in the $N_2$ phase and to 0.44% in the yeast phase (primarily due to the less efficient way of deoxygenation of $N_2$ substitution in the mixture compared to oxygen scavenging by yeast). On the other hand, SRS with FetalSenseM V2 present underestimation errors in $StO_2$ during full oxygenation (mean difference error of -11.40%, maximum difference error of -14.45%), whereas underestimation errors in $StO_2$ are identified with dual slope with FetalSenseM V2 during both deoxygenation with $N_2$ (mean difference error of -11.73%, maximum difference error of -27.06%) and with yeast (mean difference error of -19.18%, maximum difference error of -35.46%).

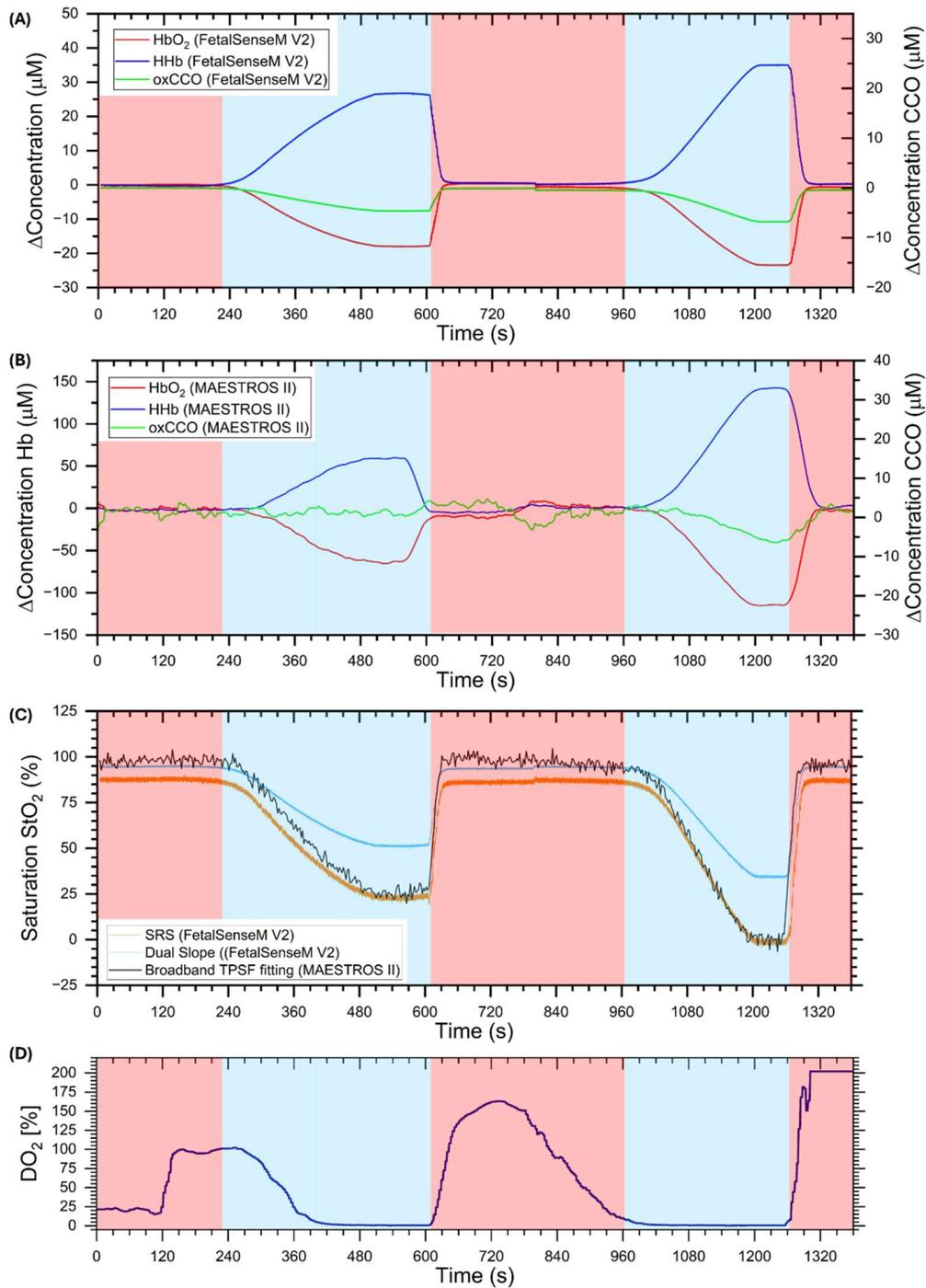

**Fig. 7** Dynamic changes in concentration of HbO$_2$ (red), HHb (blue) and oxCCO (green) over time during PROTOCOL 2, for both (A) FetalSenseM V2 and (B) MAESTROS II, at the same SDS of 3 cm. (B) Comparison of different reconstructions of StO$_2$ during PROTOCOL 2, via SRS (orange) and dual slope (cyan) for FetalSenseM V2, as well as via TPSF fitting (black) for MAESTROS II. (C) Changes in DO$_2$ during PROTOCOL 2. The red and blue blocks correspond to oxygenation and deoxygenation phases, respectively.

*3.4 Examples of application of PROTOCOL 3*

An example of application of PROTOCOL 3, presented here, consisted in measuring the multi-layered, hybrid phantom with both FetalSenseM V1.5 and V2 during a single cycle of full oxygenation via $O_2$ bubbling ($DO_2 \geq 100\%$), and deoxygenation via yeast consumption ($DO_2 \sim 0\%$). For this purpose, PROTOCOL 3 was tested for a total thickness of the solid overlayers of 1.5 cm, according again to the median value of placental depth across ultrasound scans performed on 268 healthy participants at UCLH (ad described in Sec. 2.4), as well as based on the results of the application of PROTOCOL 0 on both optical devices (as in Sec. 2.1). The optical properties and relative thicknesses of each layer that were used are the ones reported in Table 6 of Sec. 2.4. Dynamic, relative changes $\Delta[HBO_2]$, $\Delta[oxCCO]$ and $\Delta[oxCCO]$ in the concentrations of $HbO_2$, HHb and oxCCO were calculated for each dataset acquired with FetalSenseM V1.5 and V2 using the UCLn algorithm for both instruments, and the same wavelength-dependent DPF. The results of the data analysis on PROTOCOL 3 are reported in Figure 8, at all available SDS.

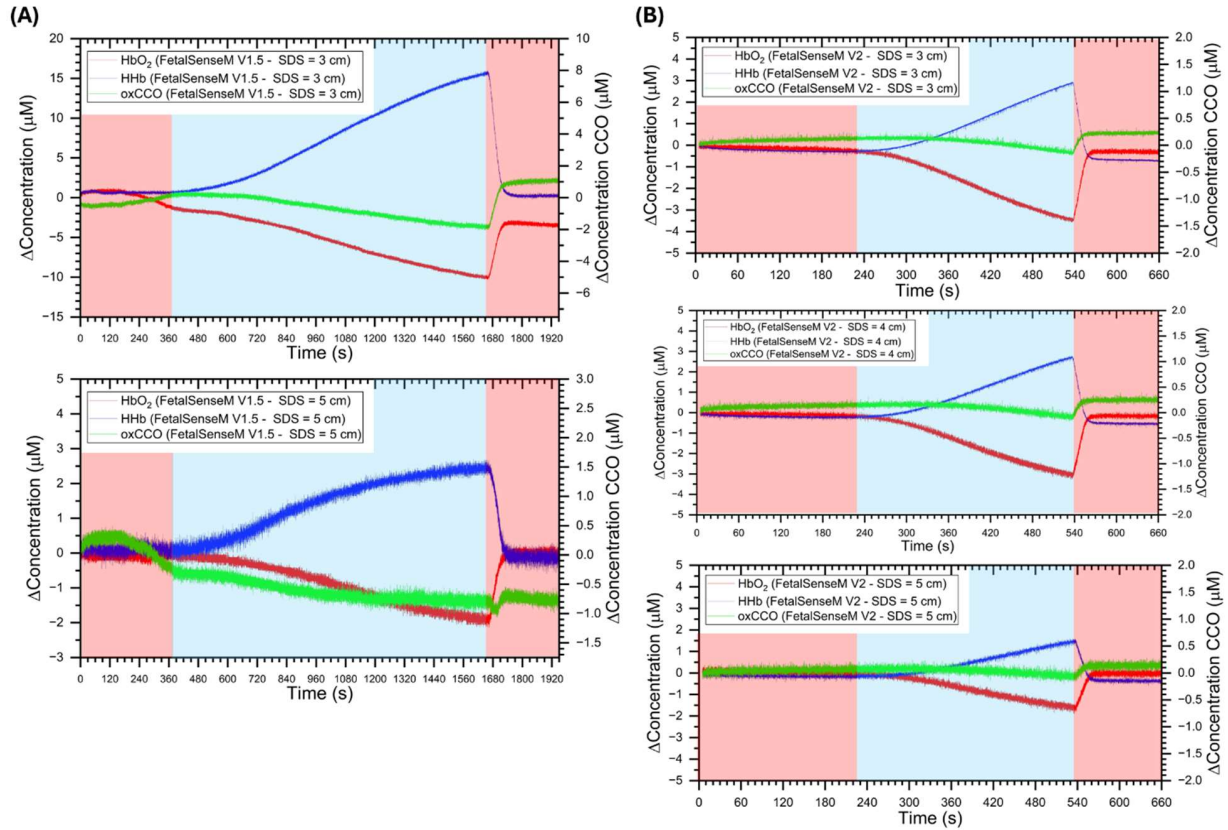

**Fig. 8** Dynamic changes in concentration of HbO$_2$ (red), HHb (blue) and oxCCO (green) over time during PROTOCOL 3, for both (A) FetalSenseM V1.5 and (B) V2, at all their available SDS (3 and 5 cm for V1.5 and 3, 4 and 5 cm for V2). The red and blue blocks correspond to oxygenation and deoxygenation phases, respectively.

The exemplary results of PROTOCOL 3 on the two tested devices offers an application case where we validated their capability of reaching a depth sensitivity of 1.5 cm for placental-like monitoring, as correct and expected dynamic changes in concentrations of all the chromophores of interest are reconstructed for both FetalSenseM V1.5 and V2 at each SDS, in the liquid phantom beneath the overlayers. Furthermore, the data provided by PROTOCOL 3 has the benefit of giving insight on the effect that the overlayers have on the quantification of the reconstructed dynamic changes in Δ[HBO$_2$], Δ[oxCCO] and Δ[oxCCO]: from Fig. 8, it can be noted that discrepancies between SDS in the magnitudes of the responses of HbO$_2$, HHb and oxCCO measured by both FetalSenseM

V1.5 and V2 are present, most significantly in the case of FetalSenseM V1.5. At 3-cm SDS, FetalSenseM V1.5 estimates changes in concentration $HbO_2$ and $HHb$ of in the order of 10 to 15 μM (absolute maximum value), compared to about 2 μM at SDS equal to 5 cm. This underestimation of the reconstructed response at longer SDS is less prominent in FetalSenseM V2, where SDS of 3 and 4 cm are much comparable in the magnitude of their changes in concentration of $HbO_2$, $HHb$ and oxCCO, with larger underestimates present at SDS equal to 5 cm (magnitude differences in the order of 1.5 to 2 μM in absolute maximum value). The latter phenomenon may be strongly related to the differences in photon pathlength between the two different overlayers against each other and also against and the liquid underneath, that are not accounted for with the use of the current, single DPF in the UCLn algorithm used for this instance. Nonetheless, the findings suggest that PROTOCOL 3 could indeed be used to recognise and quantify such types of errors, thus opening to the possibility of developing post-processing corrections of the data, or even more realistic and efficient ways to calculated DPF values for the specific application of placental monitoring and its specific geometry.

## 4   Conclusions

We have developed and presented here OptoCENTAL, a standardised platform based on various types of optical phantoms (from digital, to solid, liquid and even hybrid) for a complete characterisation and validation of any light-based instrumentation aimed at clinical placenta monitoring. The platform is based around an initial modelling and computational protocol, PROTOCOL 0, followed by three hardware-and-software-testing protocols, named PROTOCOL 1 to 3. PROTOCOL 0 is based on MC modelling of both a virtual tested instrument and a realistic, multilayered geometry and optical properties of the placenta and abdominal wall from real, hospital ultrasound scans. It thus provides a tailored simulation and theoretical validation that can even be personalised according to a specific subject or patient, due to the use of actual clinical, structural datasets.

PROTOCOL 1 assesses the hardware basic performances in terms of its key features, such as SNR, noise, linearity, stability and repeatability. It thus provides a thorough screening of the basic instrument performances that can be quantified accurately.

PROTOCOL 2 tests for the capability of any system in retrieving correct physiological monitoring of the placenta, reproducing the exact optical signatures from the major chromophores of interest in such type of tissue, including haemoglobin and CCO. It is thus extremely specific to the real placental optical contrasts, compared to other proxies, such as dyes or colorants, which never fully reproduce the exact spectra of the target biomarkers, nor are controllable in changing their optical properties according to the very specific spectral responses occurring in placental physiology.

Finally, PROTOCOL 3 validates the clinical applicability of any device in a realistic setting mimicking the placenta and its overlayers, delivering truthful evaluation of depth sensitivity in a very specific scenario that is the closest we can find in the literature to a real, clinical one, yet in a completely controllable and reproducible manner.

Overall, OptoCENTAL provides for the first time (to our knowledge) a systematic and repeatable set of bench-testing procedures for assessing any piece of photonic instrumentation with the final goal of measuring the placenta optically, eventually demonstrating their applicability in the clinics. OptoCENTAL has the main, key advantages of:

- Being quantitative, comprehensive, and realistic, as it provides controllable proxies of the placenta, its physiology, its optical properties, as well as its abdominal overlayers and their structure and contribution to the overall measurable signal.

- Being rapid (as it can be performed in up to a week of acquisitions), easy to conduct, reproducible, and relatively inexpensive, since the cost for manufacturing a single batch of solid phantoms of about 1 kg stands at less than 20£, while the liquid phantoms uses readily available, commercial material (PBS, intralipid, water, plexiglass panels for the housing and probes for monitoring and controlling the mixture), including the possibility to use commercial horse blood, with equivalent results as for human blood.

- Allowing users to flexibly apply the testing procedures to any optical instrument based on diffuse optics for placental monitoring, as well as to compare and cross-correlate the performances of different devices, spanning from CW to TD-NIRS system, wireless and wearable units, and virtually much more.

Additionally, future work on further enhancing the OptoCENTAL platform is currently ongoing: beyond the hereby proposed three protocols on the described types of physical, optical phantoms, additional designs could be implemented in the future to further enhance the testing capabilities of the procedure. For instance, even more realistic configurations than the one used currently in PROTOCOL 3 could be envisioned using 3D printing[44–47] or more advanced casting techniques to create solid, compartmentalised matrixes that replicates even the curvature of the human abdomen and the complex geometries of its inner structures, such as the uterus. These matrixes could then act as a skeleton for the phantom design, to be the filled with suitable liquid phantom solutions like the one used in PROTOCOL 2 to be dynamically controlled, as to replicate the optical properties of each single layer. This type of design would then require appropriate pumping system to circulate each solution, unlocking then the possibility to also mimic the flow of blood and other bodily fluids. However, to implement a solution like the one here proposed, it would certainly require more extensive financial resources and longer manufacturing times, thus increasing the complexity of the platform, its costs and easiness of reproducibility. Thus, analysis of the benefits against costs and accessibility to the largest possible number of users/testers should be conducted accordingly, to estimate the convenience of such upgrades and their effective utility for the purpose of enhancing realism of the optical phantoms and enhancing validation outcomes.

*Disclosures*

The authors declare that there are no financial interests, commercial affiliations, or other potential conflicts of interest that could have influenced the objectivity of this research.

*Code, Data, and Materials*

All data in support of the findings of this paper are available within the article or as supplementary material. The code for the Monte Carlo simulator and the digital phantoms used in PROTOCOL 0 of OptoCENTAL is available at: GITHUB LINK

*Acknowledgments*

LG, UH, MT, DG, FL, NRZ, SM and IL were supported by the Wellcome Trust (219610/Z/19/Z), Wellcome Leap as part of the In Utero programme, Wellcome / EPSRC Centre for Interventional and Surgical Sciences (WEISS), and National Institute for Health Research University College London Hospitals Biomedical Research Centre. FL, AA and IL are supported by: (1) the HyperProbe project, as part of the European Union's Horizon Europe research and innovation programme under grant agreement No 101071040 and of the UKRI grant number 10048387; and (2) the fastMOT project, as part of the EU's HORIZON EUROPE programme under grant agreement number 101099291 and of the UK Research and Innovation (UKRI) under the UK government's Horizon Europe funding guarantee (grant number 10063660).

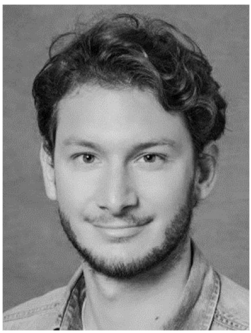

**Luca Giannoni** is a biomedical optics research fellow with the Biomedical Optics Research Laboratory, Department of Medical Physics and Biomedical Engineering at University College London. He received his PhD in Medical Imaging from University College London (UCL) in 2020. His current research interests focus on developing cutting-edge broadband near-infrared spectroscopy systems, as well as on designing and validating cost-effective, compact HSI devices for clinical translation. He is a member of SPIE.

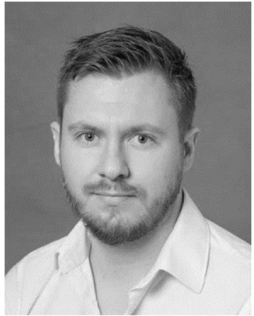

**Frédéric Lange** received his PhD from the University of Lyon and INSA de LYON in 2016. He is now a senior research associate with the Biomedical Optics Research Laboratory, Department of Medical Physics and Biomedical Engineering at University College London. His current main research interests are in the development of novel optical technologies to monitor tissue's oxygenation and metabolism, with a specific interest for non-invasive brain monitoring in healthy (i.e., brain development/neuroscience) and pathological conditions.

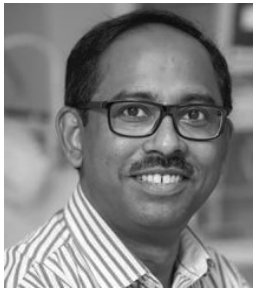

**Subhabrata Mitra** is a professor of Neonatal Medicine at University College London. He is a clinical scientist and leads the perinatal optical monitoring group, interested in developing better assessment of placental function for prediction of pregnancy outcomes and early biomarkers of perinatal brain injury, using a multimodal platform of optical neuromonitoring, integrative systematic physiology, and advanced neuroimaging.

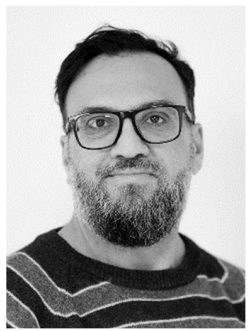

**Ilias Tachtsidis** is a professor at University College London. He leads the Multimodal Spectroscopy and MetaboLight groups in the Biomedical Optics Research Laboratory at the department of Medical Physics and Biomedical Engineering. Tachtsidis is a multidisciplinary scientist with a research portfolio encompassing engineering, physics, computing, neuroscience and clinical medicine. His research focus is technology development in optical neuroimaging (devices, algorithms), through applications (clinical, neuroscience), to data analytics towards generating clinical information and knowledge (computational models).

Biographies and photographs for the other authors are not available.

**Caption List**

**Fig. 1** Various types of optical phantoms used in PROTOCOL 1, 2 and 3 of OptoCENTAL.

**Fig. 2** The three Placental-monitoring instruments tested with OptoCENTAL.

**Fig. 3** Examples of results from PROTOCOL 0 on MAESTROS II.

**Fig. 4** Examples of results from PROTOCOL 0 on the FetalSenseM V.15 and V2.

**Fig. 5** Examples of results from PROTOCOL 1 on the FetalSenseM V.15 and V2.

**Fig. 6** Examples of results from PROTOCOL 2 on microCYRIL.

**Fig. 7** Examples of results from PROTOCOL 2 on the FetalSenseM V2 and MAESTROS II.

**Fig. 8** Examples of results from PROTOCOL 3 on the FetalSenseM V.15 and V2.

Table 1 Volume fractions of melanosome used in PROTOCOL 0.

Table 2 Volume content contributions of water and fat used in PROTOCOL 0

Table 3 Recipe for manufacturing solid, homogeneous phantoms used in PROTOCOL 1.

Table 4 Theoretical vs. measured optical properties of the three phantoms used in PROTOCOL 1.

Table 5 Compositions and main parameters of the liquid optical phantom used in PROTOCOL 2.

Table 6 Optical properties and thicknesses of the solid overlayers used in PROTOCOL3.

Table 7 Average optical properties of the overlayers of the abdominal wall above the placenta.

Table 8 Main features of the three NIRS instruments enrolled in the application of OptoCENTAL.

Table 9 Comparison of SNR assessment for FetalSenseM V1.5 and V2 during PROTOCOL 1.

Table 10 Comparison of noise assessment for FetalSenseM V1.5 and V2 during PROTOCOL 1.

Table 11 Linearity correlation coefficients for FetalSenseM V1.5 and V2 during PROTOCOL 1.

*Supplementary Materials*

Table S.1 Numerical results for example sensitivity profiles of MAESTROS II from PROTOCOL 0.

| Bin start (ns) | Bin end (ns) | Skin | Adipose | Muscle | Placenta |
|---|---|---|---|---|---|
| **Detector 1** | | | | | |
| 0.23 | 0.41 | 5.00 | 5.38 | 14.33 | 75.30 |
| 0.43 | 0.61 | 4.80 | 5.20 | 13.93 | 76.06 |
| 0.63 | 0.81 | 4.64 | 5.04 | 13.56 | 76.76 |
| 0.83 | 1.01 | 4.49 | 4.89 | 13.20 | 77.43 |
| 1.03 | 1.21 | 4.37 | 4.74 | 12.85 | 78.04 |
| 1.23 | 1.41 | 4.21 | 4.61 | 12.53 | 78.65 |
| 1.43 | 1.61 | 4.08 | 4.48 | 12.23 | 79.21 |
| 1.63 | 1.81 | 3.96 | 4.36 | 11.94 | 79.75 |
| 1.83 | 2.01 | 3.85 | 4.24 | 11.66 | 80.24 |
| 2.03 | 2.21 | 3.76 | 4.14 | 11.39 | 80.71 |
| 2.23 | 2.41 | 3.64 | 4.04 | 11.14 | 81.18 |
| 2.43 | 2.61 | 3.54 | 3.94 | 10.90 | 81.62 |
| 2.63 | 2.81 | 3.46 | 3.85 | 10.66 | 82.03 |
| **Detector 2** | | | | | |
| 0.39 | 0.57 | 4.84 | 5.22 | 13.99 | 75.95 |
| 0.59 | 0.77 | 4.66 | 5.06 | 13.61 | 76.67 |
| 0.79 | 0.97 | 4.52 | 4.90 | 13.25 | 77.33 |
| 0.99 | 1.17 | 4.37 | 4.76 | 12.91 | 77.97 |
| 1.19 | 1.37 | 4.21 | 4.62 | 12.58 | 78.59 |
| 1.39 | 1.57 | 4.09 | 4.49 | 12.27 | 79.14 |
| 1.59 | 1.77 | 3.95 | 4.37 | 11.98 | 79.70 |
| 1.79 | 1.97 | 3.85 | 4.26 | 11.70 | 80.19 |
| 1.99 | 2.17 | 3.75 | 4.15 | 11.43 | 80.67 |
| 2.19 | 2.37 | 3.65 | 4.05 | 11.17 | 81.13 |
| 2.39 | 2.57 | 3.58 | 3.95 | 10.93 | 81.54 |
| 2.59 | 2.77 | 3.46 | 3.86 | 10.70 | 81.98 |
| 2.79 | 2.97 | 3.39 | 3.77 | 10.47 | 82.37 |
| 2.99 | 3.17 | 3.31 | 3.68 | 10.26 | 82.75 |
| 3.19 | 3.37 | 3.24 | 3.59 | 10.05 | 83.12 |
| 3.39 | 3.57 | 3.17 | 3.52 | 9.85 | 83.45 |
| **Detector 3** | | | | | |
| 0.57 | 0.75 | 4.66 | 5.05 | 13.63 | 76.66 |
| 0.77 | 0.95 | 4.49 | 4.90 | 13.27 | 77.35 |
| 0.97 | 1.15 | 4.38 | 4.76 | 12.92 | 77.94 |
| 1.17 | 1.35 | 4.24 | 4.62 | 12.60 | 78.54 |
| 1.37 | 1.55 | 4.11 | 4.49 | 12.29 | 79.11 |
| 1.57 | 1.75 | 3.99 | 4.37 | 11.99 | 79.64 |
| 1.77 | 1.95 | 3.89 | 4.26 | 11.71 | 80.15 |
| 1.97 | 2.15 | 3.77 | 4.15 | 11.44 | 80.63 |
| 2.17 | 2.35 | 3.66 | 4.05 | 11.19 | 81.11 |
| 2.37 | 2.55 | 3.56 | 3.95 | 10.94 | 81.55 |
| 2.57 | 2.75 | 3.48 | 3.86 | 10.71 | 81.96 |
| 2.77 | 2.95 | 3.36 | 3.76 | 10.49 | 82.39 |
| 2.97 | 3.15 | 3.30 | 3.68 | 10.27 | 82.75 |
| 3.17 | 3.35 | 3.24 | 3.60 | 10.06 | 83.09 |
| 3.37 | 3.55 | 3.17 | 3.52 | 9.87 | 83.44 |
| 3.57 | 3.75 | 3.07 | 3.45 | 9.68 | 83.80 |
| 3.77 | 3.95 | 3.00 | 3.37 | 9.49 | 84.13 |

| Detector 4 | | | | | |
|---|---|---|---|---|---|
| Bin start (ns) | Bin end (ns) | Skin | Adipose | Muscle | Placenta |
| 0.59 | 0.77 | 4.62 | 5.00 | 13.56 | 76.82 |
| 0.79 | 0.97 | 4.45 | 4.86 | 13.20 | 77.49 |
| 0.99 | 1.17 | 4.34 | 4.72 | 12.86 | 78.08 |
| 1.19 | 1.37 | 4.19 | 4.59 | 12.54 | 78.68 |
| 1.39 | 1.57 | 4.06 | 4.46 | 12.23 | 79.25 |
| 1.59 | 1.77 | 3.94 | 4.34 | 11.94 | 79.78 |
| 1.79 | 1.97 | 3.84 | 4.23 | 11.66 | 80.27 |
| 1.99 | 2.17 | 3.74 | 4.12 | 11.40 | 80.74 |
| 2.19 | 2.37 | 3.63 | 4.02 | 11.15 | 81.20 |
| 2.39 | 2.57 | 3.54 | 3.92 | 10.90 | 81.64 |
| 2.59 | 2.77 | 3.45 | 3.83 | 10.67 | 82.05 |
| 2.79 | 2.97 | 3.36 | 3.74 | 10.45 | 82.45 |
| 2.99 | 3.17 | 3.29 | 3.66 | 10.24 | 82.81 |
| 3.19 | 3.37 | 3.22 | 3.58 | 10.03 | 83.16 |
| 3.39 | 3.57 | 3.14 | 3.51 | 9.84 | 83.51 |
| 3.59 | 3.77 | 3.07 | 3.44 | 9.65 | 83.84 |
| 3.79 | 3.97 | 2.99 | 3.37 | 9.47 | 84.17 |
| 3.99 | 4.17 | 2.94 | 3.29 | 9.29 | 84.48 |
| 4.19 | 4.37 | 2.88 | 3.23 | 9.12 | 84.77 |

Table S.2 Numerical results for example sensitivity profiles of the FetalSenseM V1.5 from PROTOCOL 0.

| Detector | Skin | Adipose | Muscle | Placenta |
|---|---|---|---|---|
| 3cm | 8.70 | 24.90 | 49.61 | 16.79 |
| 5cm | 5.28 | 16.99 | 45.95 | 31.78 |

**Table S.3** Numerical results for example sensitivity profiles of the FetalSenseM V2 from PROTOCOL 0.

| Detector | Skin | Adipose | Muscle | Placenta |
|---|---|---|---|---|
| 3cm | 4.04 | 7.54 | 9.54 | 54.93 |
| 4cm | 3.04 | 6.08 | 8.27 | 62.45 |
| 5cm | 2.43 | 4.73 | 6.80 | 69.43 |